\documentclass[aps,prl,reprint,groupedaddress]{revtex4-1}

\pdfoutput=1

\usepackage{amsmath}
\usepackage{amsfonts}
\usepackage{graphicx}
\usepackage{newtxtext,newtxmath}
\usepackage{comment}
\usepackage{dcolumn}
\usepackage{natbib}
\usepackage{xcolor}

\definecolor{darkslateblue}{HTML}{2E30A0}

 \usepackage{hyperref}
 \hypersetup{
    colorlinks=true,       
    linkcolor=darkslateblue,          
    citecolor=darkslateblue,        
    filecolor=magenta,      
    urlcolor=darkslateblue           
}

\renewcommand{\vec}[1]{\boldsymbol{\mathbf{#1}}}

\newcolumntype{d}[1]{D{.}{.}{#1}}

\let\originalleft\left
\let\originalright\right
\renewcommand{\left}{\mathopen{}\mathclose\bgroup\originalleft}
\renewcommand{\right}{\aftergroup\egroup\originalright}

\renewcommand{\epsilon}{\varepsilon}
\newcommand{\eb}{\ensuremath{\epsilon_b}}
\renewcommand{\vec}[1]{\boldsymbol{\mathbf{#1}}}

\begin{document}

\frenchspacing

\title{Positron Binding and Annihilation in Alkane Molecules}
\author{A.~R. Swann}
\email{a.swann@qub.ac.uk}
\author{G.~F. Gribakin}
\email{g.gribakin@qub.ac.uk}
\affiliation{
School of Mathematics and Physics, Queen's University Belfast, University Road, Belfast BT7 1NN, United Kingdom}
\date{\today}

\begin{abstract}
A model-potential approach has been developed to study positron interactions with molecules. Binding energies and annihilation rates are calculated for positron bound states with a range of alkane molecules, including rings and isomers. The calculated binding energies are in good agreement with experimental data, and the existence of a second bound state for $n$-alkanes (C$_n$H$_{2n+2}$) with $n\geq12$ is predicted in accord with experiment. The annihilation rate for the ground positron bound state scales linearly with the square root of the binding energy.
\end{abstract}

\maketitle

The ability of positrons to bind to molecules underpins the spectacular phenomenon of resonantly enhanced positron annihilation observed in most polyatomic gases \cite{Gribakin10}. In this process, the positron is captured by the molecule, its excess energy transferred into molecular vibrations \cite{Gribakin00,Gribakin01}. The corresponding annihilation rates depend strongly on the molecular size and display remarkable chemical sensitivity \cite{Deutsch51,Paul63,Heyland82,Surko88,Iwata95}. Observation of energy-resolved resonant annihilation \cite{Gilbert02} has also enabled measurements of the positron binding energy $\eb$. Binding energies ranging from few to few hundred of meV, have been determined experimentally for over 70, mostly nonpolar, molecular species, including alkanes, aromatics, partially halogenated hydrocarbons, alcohols, formates, and acetates \cite{Danielson09,Danielson10,Danielson12,Danielson12a}.

This body of data is barely understood from a theoretical standpoint, in spite of a long history of the question \cite{Schrader10,Tachikawa01}. Nearly all existing calculations of positron binding considered strongly polar molecules, where binding is guaranteed at any level of theory \footnote{A static molecule with a dipole moment greater than 1.625 debye will always bind an electron or positron \cite{Crawford67}, while for a molecule that is free to rotate, the critical value dipole moment increases with the angular momentum of the molecule \cite{Garrett71}.}. A variety of methods were used, including Hartree-Fock \cite{Kurtz81}, configuration interaction \cite{Strasburger96,Chojnacki06,Gianturco06}, diffusion Monte Carlo \cite{Bressanini98,Mella00,Kita09}, explicitly correlated Gaussians \cite{Bubin04}, and the any-particle molecular-orbital approach \cite{Romero14}. The majority of calculations examined simple diatomics, such as alkali hydrides \cite{Mella00} and metal oxides \cite{Bressanini98}, or triatomics: hydrogen cyanide \cite{Chojnacki06,Kita09} and CXY (X, Y = O, S, Se) \cite{Koyanagi13} (see Ref.~\cite{Swann18} for more information). In spite of this effort, of all the molecules studied experimentally, theoretical predictions are available only for five strongly polar species (acetaldehyde, propanal, acetone, acetonitrile, and propionitrile \cite{Tachikawa11,Tachikawa12,Tachikawa14}), and the best agreement does not exceed 25\% (for acetonitrile, $\eb =136$~meV, theory \cite{Tachikawa14}, vs. 180~meV, experiment \cite{Danielson10}). Critically, quantum-chemistry calculations have so far failed to predict positron binding to nonpolar molecules with any degree of accuracy \footnote{Calculations for CS$_2$ \cite{Koyanagi13} gave a negative binding energy, vs. measured $\eb = 75$~meV \cite{Danielson10}. Scattering calculations for allene indicate binding \cite{Barbosa17}, but no estimate of $\eb$ is provided.}.

To address this problem, we construct a simple physical model that enables calculations of positron binding to a wide range of polyatomic species and has predictive capability. 
We apply the model to a range of alkanes and find good agreement with experiment, which confirms that the effective positron-molecule potential is largely ``additive'' and distributed over the molecule, and that its short-range part is just as important as the long-range behavior determined by the molecular polarizability.
While this short-range part cannot be described \textit{ab initio} with the required accuracy, we show that it can be parametrized in a  reliable way. 
This opens the way for calculating positron binding energies, annihilation rates, and $\gamma$ spectra for all molecules that have been studied experimentally and for making predictions for other molecules. Understanding positron binding to molecules also sheds light on its counterpart---the problem of electron attachment to molecules and formation of molecular anions.

\textit{Theoretical approach.}---Since accurate predictions of positron binding to polyatomic molecules are
beyond the capacity of the best \textit{ab initio} calculations, we use a model-correlation-potential approach \cite{Swann18}. The electrostatic potential $V_\text{st}$ of the molecule is calculated at the Hartree-Fock level using the standard 6-311++G($d$,$p$) basis, and then a potential that describes long-range polarization of the molecular electron cloud by the positron is added. The explicit form of this potential is 
\begin{equation}
V_\text{cor}(\vec{r}) = -  \sum_A \frac{\alpha_A}{2 \lvert \vec{r}-\vec{r}_A\rvert^{4}} \left[ 1 - \exp\left( - \lvert \vec{r}-\vec{r}_A\rvert^6/\rho_A^6\right)\right] ,
\end{equation}
 where the sum is over the  atoms $A$ in the molecule, $\vec{r}$ is the position of the positron, and $\vec{r}_A$ is the position of nucleus $A$, relative to an arbitrary origin.
(Atomic units (a.u.) are used throughout, unless stated otherwise.)
This model potential uses the hybrid polarizabilities $\alpha_A$ of the molecule's constituent atoms \cite{Miller90}, which take into account the chemical environment of the atom within the molecule. The factor in  brackets provides a short-range cutoff, characterized by the cutoff radius $\rho _A$, which is a free parameter of the theory. 
Its values are expected to be comparable to the radii of the atoms involved, e.g., in the range  of 1--3~a.u. 
Far from the molecule, the potential takes the asymptotic form $V_\text{cor}(\vec{r})\simeq-\alpha/2r^4$, where $\alpha=\sum_A\alpha_A$ is the  molecular polarizability \footnote{The present implementation regards atomic contributions to $V_\text{cor}(\vec{r})$ as spherically symmetric. In future, they can be made anisotropic, to correlate with the directions of adjacent bonds.}. 
The short-range part of the potential accounts for other important electron-positron correlation effects, such as virtual positronium formation.

The Schr\"odinger equation for the total potential $V_\text{st} + V_\text{cor}$ is solved to obtain the positron binding energy $\eb$ and the positron wave function.
In practice, this is done using the standard quantum-chemistry package \textsc{gamess} with the \textsc{neo} plugin \cite{Schmidt93,Gordon05,Webb02,Adamson08}, which we have modified to include the model potential $V_\text{cor}$ \cite{Swann18}. 
We use an even-tempered Gaussian basis consisting of 12 $s$-type  primitives centered on each C nucleus, with exponents $0.0001\times3^{i-1}$ ($i=1$--12), and eight $s$-type  primitives centered on each H nucleus, with exponents $0.0081\times3^{i-1}$ ($i=1$--8).

\textit{Binding energies for alkanes.}---Here we apply the method to alkanes, which are nonpolar or very weakly polar molecules. While no quantum-chemistry calculations of positron binding have been reported for them before, positron binding energies have been measured for most of the  $n$-alkanes C$_n$H$_{2n+2}$ with $n=3$--16 (methane CH$_4$ does  not support a positron bound state, and while ethane C$_2$H$_6$ appears to bind a positron, $\eb$ is too small to measure), and also for isopentane C$_5$H$_{12}$, cyclopropane C$_3$H$_6$, and cyclohexane C$_6$H$_{12}$ \cite{Young08}. The binding energy for the $n$-alkanes was found to increase close to linearly with $n$, and a second bound state was observed for $n\geq 12$. 

\begin{figure}
\includegraphics[width=0.9\columnwidth]{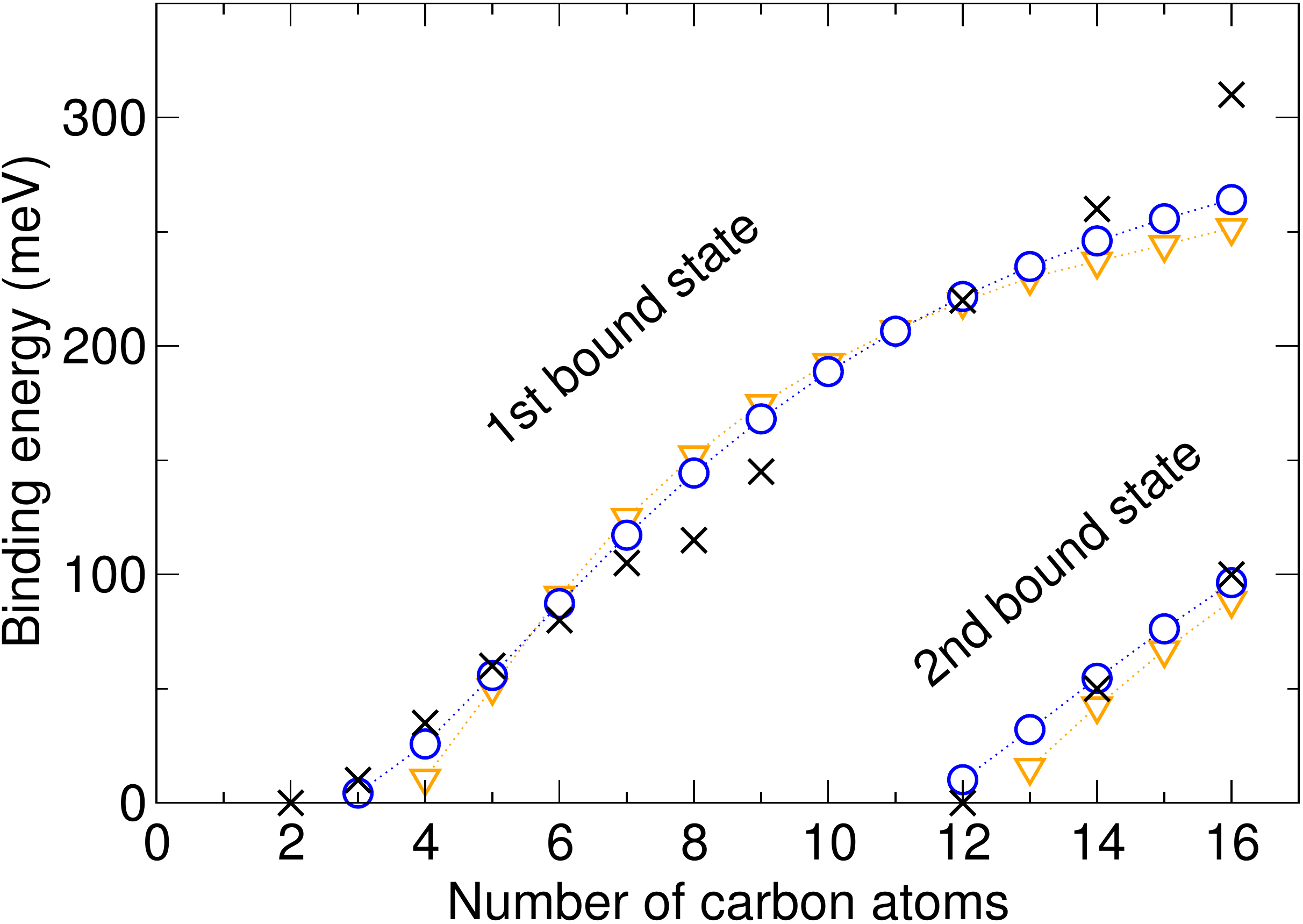}
\caption{\label{fig:alkanes}Positron binding energies for $n$-alkane molecules C$_n$H$_{2n+2}$. Black crosses, experiment \cite{Young08}; blue circles, present calculation; orange triangles, zero-range potential model \cite{Gribakin09}.}
\end{figure}

We choose $\alpha _\text{C}=7.096$ and $\alpha _\text{H}=2.650$~a.u., which provide the best fit, $\alpha =n\alpha _\text{C}+(2n+2)\alpha _\text{H}$, of the polarizabilities of alkanes \cite{Miller90}. We use the same cutoff radius for the C and H atoms, and set $\rho _A=2.25$~a.u. to reproduce the measured $\eb =220$~meV for dodecane C$_{12}$H$_{26}$. Figure~\ref{fig:alkanes} shows the values of $\eb$ obtained for the $n$-alkanes C$_n$H$_{2n+2}$ in terms of $n$. 
Also shown are the experimental data \cite{Young08} and the crude zero-range-potential (ZRP) calculations (in which each of the CH$_2$ or CH$_3$ groups was replaced by a short-range deltalike potential, whose strength was chosen to fit the binding energy for dodecane) \cite{Gribakin09}.
The present calculations and the experimental data are also shown in Table \ref{tab}.
\begin{table}
\caption{\label{tab}Calculated binding energies $\eb$, independent-particle-approximation contact densities $\delta_{ep}^{(0)}$, and enhanced and renormalized contact densities $\delta_{ep}$ for  $n$-alkane molecules C$_n$H$_{2n+2}$. Also shown are the experimental (exp.) binding energies \cite{Young08}. Square brackets indicate powers of 10.}
\begin{ruledtabular}
\begin{tabular}{ld{3}ccc}
$n$ & \multicolumn{1}{c}{$\eb$} & \multicolumn{1}{c}{$\eb$ (exp.)} & \multicolumn{1}{c}{$\delta_{ep}^{(0)}$} & \multicolumn{1}{c}{$\delta_{ep}$} \\
& \multicolumn{1}{c}{\text{(meV)}} & (meV) & (a.u.) & (a.u.) \\
\hline
2 & -2.177\footnotemark[1] & {>}0 &  $-$ & $-$ \\
3 & 4.302 & 10 & $5.717[-4]$ & $2.605[-3]$  \\
4 & 25.81 & 35 & $1.600[-3]$ & $7.221[-3]$ \\
5 & 55.75 & 60 & $2.547[-3]$ & $1.138[-2]$ \\
6 & 87.23 & 80 & $3.328[-3]$ & $1.476[-2]$ \\
7 & 117.2 & 105 & $3.948[-3]$ & $1.740[-2]$ \\
8 & 144.4 & 115 & $4.426[-3]$ & $1.943[-2]$ \\
9 & 168.1 & 145 & $4.791[-3]$ & $2.096[-2]$ \\
10 & 188.8 & & $5.068[-3]$ & $2.209[-2]$ \\
11 & 206.5 & & $5.280[-3]$ & $2.300[-2]$ \\
12 & 221.7 & 220 & $5.445[-3]$ & $2.367[-2]$ \\
12\footnotemark[2] & 10.14 & 0 & $2.587[-3]$ & $1.158[-2]$  \\
13 & 234.8 & & $5.570[-3]$ & $2.420[-2]$ \\
13\footnotemark[2] & 32.12 & & $3.337[-3]$ & $1.482[-2]$ \\
14 & 246.0 & 260 & $5.666[-3]$ & $2.458[-2]$ \\
14\footnotemark[2] & 54.56 & 50 & $3.858[-3]$ & $1.703[-2]$ \\
15 & 255.7 & & $5.744[-3]$ & $2.492[-2]$ \\
15\footnotemark[2] & 76.19 & & $4.257[-3]$ & $1.872[-2]$ \\
16 & 264.1 & 310 & $5.805[-3]$ & $2.516[-2]$ \\
16\footnotemark[2] & 96.40 & 100 & $4.568[-3]$ & $2.004[-2]$
\end{tabular}
\end{ruledtabular}
\footnotetext[1]{With no binding, this value is determined by the size of the basis.}
\footnotetext[2]{Second bound state.}
\end{table}
We obtain generally  very good agreement with the experimental data. For  $n=3$--7, our results follow the near-linear trend of the experiment much more closely than do the zero-range-potential calculations. In particular, we report a positive binding energy for $n=3$ (propane), where the ZRP model shows no binding. Also, the present calculation predicts the emergence of a second bound state for $n=12$ (dodecane), in agreement with experiment, while the ZRP model only shows this for $n=13$. For $n=8$ (octane) and 9 (nonane), we observe a somewhat larger discrepancy with the measured binding energies. We note, however, that the experimental data for these molecules lie slightly below the linear trend set by the other molecules. This difference may therefore be due to an experimental error. From $n\approx 12$ upwards, the calculated binding energies show signs of saturation and drop below the near-linear trend observed for smaller $n$; this effect is even more pronounced in the ZRP data. Indeed,  for $n=14$ and 16, our $\eb$ for the first bound states underestimate the experimental values by 5 and 15\%, respectively, although the second bound state is still very well described.
The exact reasons for this discrepancy are not clear.
One possibility is that at room temperatures such large chain molecules may favor conformations other than linear \cite{Thomas06}, for which the calculations were performed.
At the other end of the scale, our calculations with $\rho _A=2.25$~a.u. fail to predict a bound state for $n=2$ (ethane), and it would be necessary to reduce the value of the cutoff radius to 2.09~a.u. for a bound state to appear. This likely reflects the fact that the cutoff radius can have a weak dependence on the size of the molecule, which becomes more obvious for smaller species.

\begin{figure}
\centering
\includegraphics[width=.9\columnwidth]{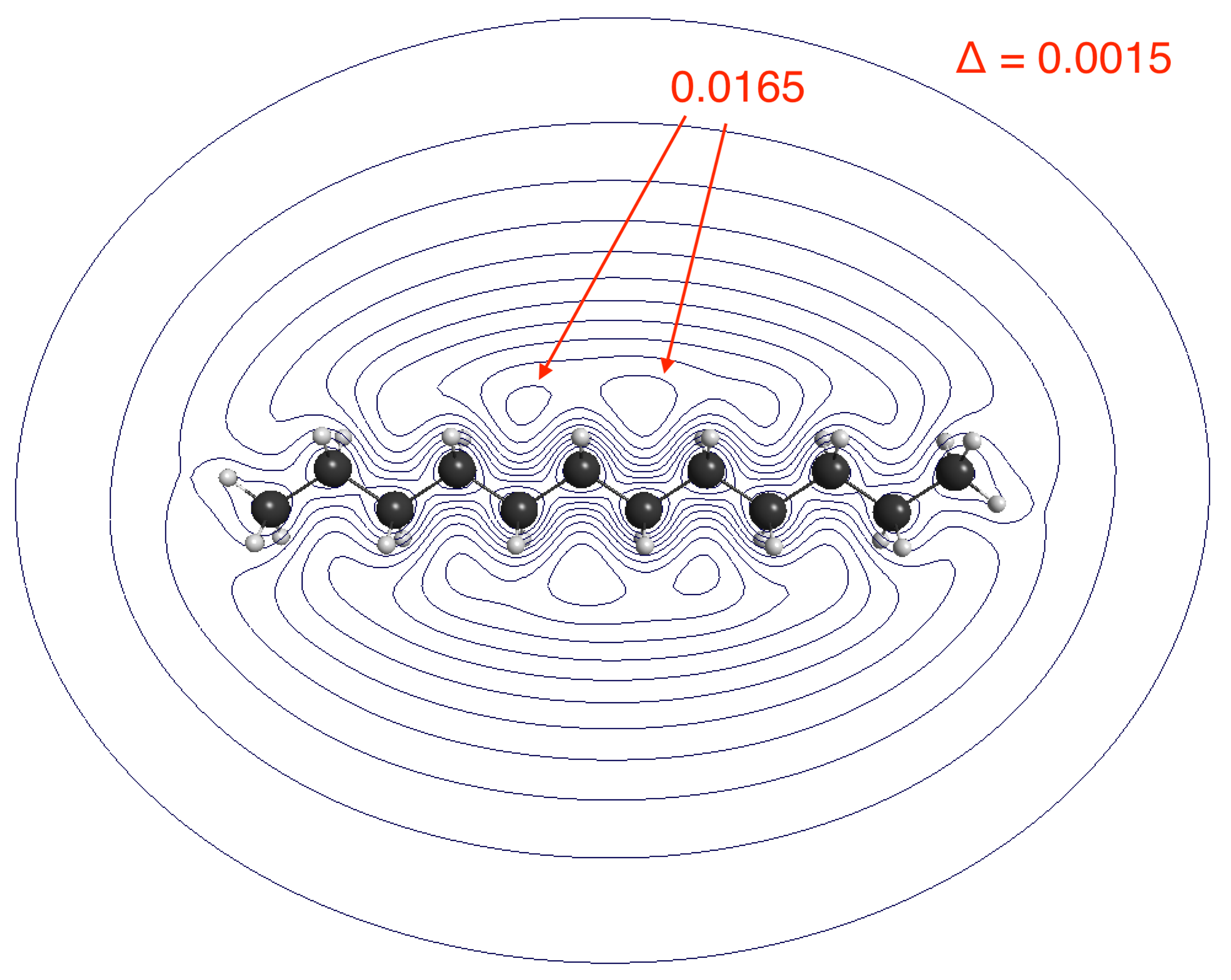} \\[1em]
\includegraphics[width=.9\columnwidth]{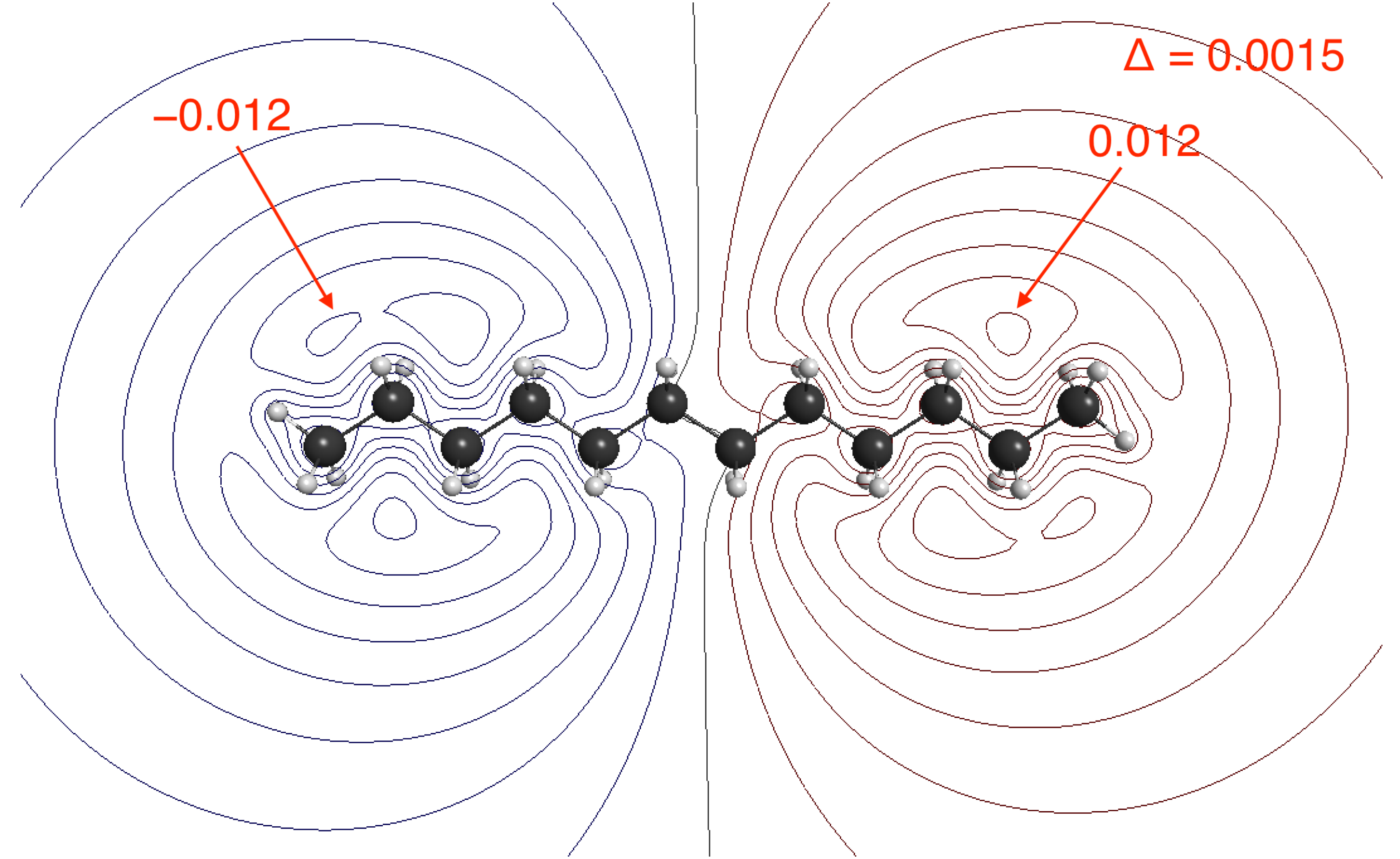}
\caption{\label{fig:3D_orbitals}Contour plots of the first (upper panel) and second (lower panel) bound positron states for dodecane C$_{12}$H$_{26}$. The contour for which the magnitude of the wave function is largest is indicated. The change in the magnitude of the wave function between neighboring contours is $\Delta=0.0015$.
} 
\end{figure}

Figure~\ref{fig:3D_orbitals} shows the shapes of the first and second bound positron orbitals for dodecane. 
We see that the positron cloud surrounds the entire molecule,
as was inferred from the analysis of measured annihilation $\gamma$-ray spectra \cite{Iwata97}.
This is in contrast to strongly polar molecules, where the bound positron is strongly localized around the negative end of the dipole \cite{Danielson12,Swann18}. The wave function of the second bound state has a $p$-wave character. It changes sign when
crossing a nodal surface (``plane'') near the centre of the molecule.

Besides the near-linear increase of the binding energy for $n$-alkanes, the experiment found that isopentane C$_5$H$_{12}$, cyclopropane C$_3$H$_6$, and cyclohexane C$_6$H$_{12}$ have the same binding energies as the $n$-alkanes with the same number of carbon atoms \cite{Young08}. Using our method, we find that the binding energy for isopentane is $\eb= 59$~meV, which is only 5\% greater than the calculated value of 56~meV for $n$-pentane. Both values are close to the experimental value $\eb= 60$~meV \cite{Young08}. (The accuracy of the experimental determination of $\eb $ is likely no better than 5~meV, due to uncertainties in the energy of the positron beam.) For neopentane, our calculations yield $\eb = 57$~meV, though there are no measurements for this isomer.
The similarity between the binding energies for the three isomers suggests the long-range behavior of $V_\text{cor}$ (which is the same in all three cases) is more important for positron binding than the effects of the molecular geometry.
The calculated values for cyclopropane and $n$-propane are $\eb=0.66$ and $4.3$~meV, respectively, while the experimental value is 10~meV. The smaller calculated binding energy for cyclopropane is due to the fact that its polarizability is 12\% smaller than that of $n$-propane. Similarly, the calculated binding energies for cyclohexane and $n$-hexane are 76 and 87~meV, respectively, which can be attributed to the 7\% smaller polarizability of cyclohexane. Experimentally, they were reported to have same binding energy of 80~meV \cite{Young08}. However, updated analysis using a somewhat higher resolution beam indicates $\eb=80$~meV for cyclohexane and $\eb=95$~meV for $n$-hexane \footnote{J. R. Danielson, S. Ghosh, and C. M. Surko (private communication).}, in close accord with the calculations.

\textit{Annihilation rates for alkanes.}---The $2\gamma$ annihilation rate for the positron from the bound state, averaged over the electron and positron spins, is given by  $\Gamma=\pi r_0^2 c \delta_{ep}$, where $r_0$ is the classical electron radius, $c$ is the speed of light, and $\delta_{ep}$ is the electron-positron \textit{contact density} in the bound state \cite{Gribakin10}. 
A useful conversion from the contact density to the annihilation rate  is $\Gamma [\text{ns}^{-1}]=50.470 \times \delta_{ep}[\text{a.u.}]$. The lifetime of the positron-molecule complex with respect to annihilation is $1/\Gamma$. 

We use the wave functions of the electronic molecular orbitals along with the positron wave function to calculate the electron-positron  contact density $\delta_{ep}$, viz.,
\begin{equation}
\delta_{ep} = \sum_i \gamma_i \int \lvert \varphi_i(\vec{r})\rvert^2 \lvert \psi(\vec{r})\rvert^2\, d\tau,
\end{equation}
where the sum is over all of the occupied Hartree-Fock electronic spin orbitals with wave functions $\varphi_i$, $\psi$ is the positron wave function, and $\gamma_i$ is an annihilation vertex \textit{enhancement factor}, specific to spin orbital $i$. The enhancement factor is introduced to improve on the independent-particle approximation by accounting for an increase of the electron density at the positron due to their Coulomb interaction \cite{Green15}. 
Similar enhancement factors are used in calculations of positron annihilation in solids \cite{Puska94,Alatalo96}.
Recent many-body-theory calculations for atoms have shown that the enhancement factors are, to a good approximation, functions of the spin-orbital energy $\epsilon_i$ \cite{Green15,Swann18}:
\begin{equation}\label{eq:enh_fac}
\gamma_i = 1 + \sqrt{\frac{1.31}{-\epsilon_i}} + \left( \frac{0.834}{-\epsilon_i} \right)^{2.15}.
\end{equation}
 We also renormalize the positron wave function, to take into account the underlying many-body nature of $V_\text{cor}$.
 The true correlation potential that describes the interaction of a positron with a many-electron system is a nonlocal and energy-dependent operator $\Sigma_E(\vec{r},\vec{r}')$ \cite{Gribakin04,Green14}. When using it in the Schr\"odinger-like Dyson equation, the negative-energy eigenvalue $\epsilon_0=-\epsilon_b$ that corresponds to a bound state becomes a function of $E$, i.e., $\epsilon_0=\epsilon_0(E)$ and has to be found self-consistently. The corresponding positron wave function is, in fact, a \textit{quasiparticle} wave function, normalized as \cite{Chernysheva88,Ludlow10}
 \begin{equation}\label{eq:a_fac}
 \int \lvert \psi(\vec{r}) \rvert^2 \, d\tau = \left(1 - \frac{\partial\epsilon_0}{\partial E} \right)^{-1} \equiv a < 1.
 \end{equation}
By considering the dependence of the binding energy on the molecular polarizability, we have determined values of $a$ for each molecule.
The values  range from $a=0.992$ for C$_3$H$_8$ to 0.933 for C$_{16}$H$_{34}$, for the first bound state, and from $a=0.967$ for C$_{12}$H$_{26}$ to 0.946 for C$_{16}$H$_{34}$, for the second bound state.

\begin{figure}
\includegraphics[width=0.9\columnwidth]{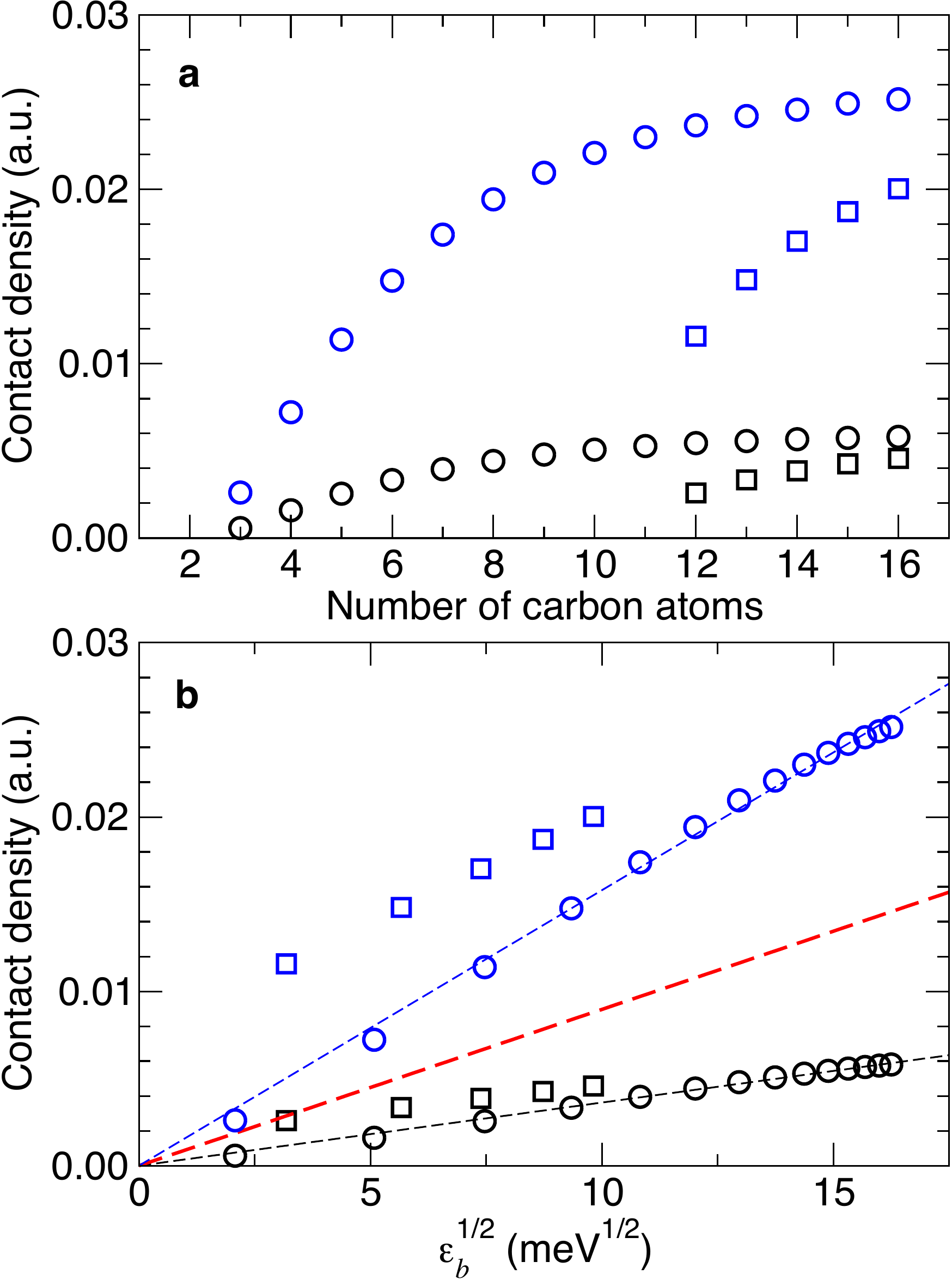}
\caption{\label{fig:ann_rates}Electron-positron contact density  for $n$-alkane molecules C$_n$H$_{2n+2}$. Panel (a) shows the contact density in terms of $n$, while panel (b) shows it as a function of $\sqrt{\eb}$. Black symbols, independent-particle approximation; blue symbols, with enhancement factors and renormalisation. Circles, first bound state; squares, second bound state. In (b), thin black and blue dashed lines are
fits of the respective first-bound-state data, and thick red dashed line is a fit of the calculated contact densities for positron-atom bound states \cite{Gribakin10}.}
\end{figure}

Figure~\ref{fig:ann_rates}(a) shows the contact density for each of the $n$-alkanes, for the first and second bound states, when the latter exists. Results are shown for the independent-particle approximation ($\gamma_i=1$ and $a=1$), and also with enhancement and renormalization, i.e., using Eqs.~(\ref{eq:enh_fac}) and (\ref{eq:a_fac}).
These data are also shown in Table \ref{tab}.
Including the enhancement factors and renormalization increases the contact density by a factor of approximately 4.5 compared to the independent-particle approximation, irrespective of the size of the molecule. The growth of the contact density with the size of the molecule is related to an increase in the positron binding energy. Previous studies of positron-atom bound states found that the contact density grew linearly with $\sqrt{\eb}$, specifically, as $\delta_{ep}\approx 9.0\times10^{-4}\sqrt{\eb}$, where $\delta_{ep}$ is  in a.u.\ and $\eb $ is in meV \cite{Gribakin01,Gribakin10}. This dependence is related to the probability of finding the positron in the vicinity of the target for weakly bound $s$-type states. Figure~\ref{fig:ann_rates}(b) shows that the contact density for the $n$-alkanes also scales linearly with $\sqrt{\eb}$, with $\delta_{ep}^{(0)}\approx 3.63\times10^{-4}\sqrt{\eb}$ in the independent-particle approximation (thin black dashed line), and $\delta_{ep}\approx 1.58\times10^{-3}\sqrt{\eb}$, when the enhancement factors and renormalization are included (thin blue dashed line).
Thus we see that the contact densities for positron bound states with alkanes are about 1.8 times greater than those for the positron-atom bound states, for the same binding energy. This difference must be related to the fact that in atoms, positron access of high-electron-density regions is always impeded by the nuclear repulsion, while in molecules it is easier for the positron to approach the electrons as they are shared between the constituent atoms. It is also worth noting that the contact density for the second bound state remains finite when its binding energy goes to zero. Such behavior is characteristic of $p$-type states that remain localized in the limit $\eb \rightarrow 0$.

We have also calculated contact densities for the isomers of pentane, cyclopropane, and cyclohexane. The values that include the enhancement factors and renormalization are $1.3\times 10^{-3}$ for cyclopropane, $1.2\times 10^{-2}$ for isopentane and neopentane, 
and $1.4\times 10^{-2}$ a.u. for cyclohexane. With the exception of cyclopropane, the contact densities for the various isomers and ring forms are very close to those for the corresponding $n$-alkane in Table \ref{tab}. For cyclopropane, the contact density is half that of $n$-propane. This is related to the fact that the calculated binding energy for cyclopropane is six times smaller than that of $n$-propane.

\textit{Summary.}---We have developed a method for calculating positron-molecule binding energies and annihilation rates and demonstrated its predictive capabilities for the alkanes. These quantities are key to understanding positron resonant annihilation in molecules. Our method allows one to investigate positron binding to other molecules that have been studied experimentally. It can also be used to make predictions for other molecular species, to guide future experimental effort and provide comparisons for more sophisticated quantum-chemistry calculations. The positron wave function can also be used to calculate the annihilation $\gamma$ spectra, where much of the experimental data \cite{Iwata97} still awaits theoretical analysis \cite{Ikabata18}.

\begin{acknowledgments}
\textit{Acknowledgments.}---We are very grateful to J. R. Danielson, S. Ghosh, and C. M. Surko for providing recent unpublished experimental data.
This work has been supported by the EPSRC UK, Grant No. EP/R006431/1. 
\end{acknowledgments}

\end{document}